\documentclass[%
 nofootinbib,
 amsmath,amssymb,
 aps, prl,twocolumn, superscriptaddress,
]{revtex4-2}
\usepackage{newtxtext,newtxmath}
\usepackage{graphicx}
\usepackage{dcolumn}
\usepackage{bm}
\usepackage{float}
\usepackage{multirow}
\usepackage[colorlinks,linkcolor=blue,urlcolor=blue,anchorcolor=blue,citecolor=blue]{hyperref}
\usepackage[capitalize]{cleveref}
\usepackage{color}
\usepackage[normalem]{ulem}

\def\nuc#1#2{\relax\ifmmode{}^{#1}{\protect\text{#2}}\else${}^{#1}$#2\fi}

\begin{document}

\title{Constraining Neutron Capture Cross Sections for \nuc{88}{Y} with Gamma-ray Strength Function in $(p,p^\prime\gamma)$ Surrogate Reaction}

\author{S. T. Zhang}
\affiliation{School of nuclear Science and Technology, University of South China, Hengyang 421001, China}

\author{W. Luo}
\thanks{Corresponding author: wenluo-ok@163.com}
\affiliation{School of nuclear Science and Technology, University of South China, Hengyang 421001, China}
\affiliation{Key Laboratory of Advanced Nuclear Energy Design and Safety, Ministry of Education, Hengyang 421001, China}

\author{D. Y. Pang}
\thanks{Corresponding author: dypang@buaa.edu.cn}
\affiliation{School of Physics, Beihang University, Beijing 100191, China}

\author{Z. C. Li}
\affiliation{School of nuclear Science and Technology, University of South China, Hengyang 421001, China}
\affiliation{Key Laboratory of Advanced Nuclear Energy Design and Safety, Ministry of Education, Hengyang 421001, China}

\author{J. Feng}
\thanks{Corresponding author: fengtyl@sohu.com}
\affiliation{Key Laboratory of Nuclear Data, China Institute of Atomic Energy, Beijing 102413, China}

\author{B. Jiang}
\affiliation{School of nuclear Science and Technology, University of South China, Hengyang 421001, China}
\affiliation{Key Laboratory of Advanced Nuclear Energy Design and Safety, Ministry of Education, Hengyang 421001, China}

\author{X. X. Li}
\affiliation{School of nuclear Science and Technology, University of South China, Hengyang 421001, China}
\affiliation{Key Laboratory of Advanced Nuclear Energy Design and Safety, Ministry of Education, Hengyang 421001, China}

\author{Y. Xu}
\affiliation{Extreme Light Infrastructure - Nuclear Physics (ELI-NP), Horia Hulubei National Institute for R\&D in Physics and Nuclear Engineering (IFIN-HH), Strada Reactorului 30, Buchurest-Magurele, 077125 Ilfov, Romania}

\author{B. H. Sun}
\affiliation{School of Physics, Beihang University, Beijing 100191, China}

\date{\today}

\begin{abstract}

We demonstrate to extract \nuc{88}{Y}$(n,\gamma)$ cross sections using the $(p,p'\gamma)$ surrogate reaction with proper treatment of the spin-parity distribution of the compound nucleus \nuc{89}{Y}. Experimental data of both $\gamma$-decay probability and $\gamma$-ray strength function are used to constrain the nuclear model parameters within a computational framework combining the Bayesian optimization and Markov chain Monte Carlo method, which helps to significantly reduce the $(n,\gamma)$ data uncertainty. The $^{88}\mathrm{Y}(n,\gamma)$ cross sections are then extracted with a narrow uncertainty of 7.6\%-23.1\% within neutron energy range of 0.01 to 3.0 MeV for the first time, where no experimental data are available. Moreover, our method is verified with the \nuc{88}{Sr}($p,\gamma$) reaction, of which the measured data are available for comparison. This work opens interesting perspectives on the matter of extracting ($n,\gamma$) reaction cross sections on unstable nuclei as surrogate reaction experiments are becoming widely available.
\end{abstract}

\maketitle

Cross sections of neutron capture reactions on unstable nuclei are pivotal to our understanding of synthesis of elements via the astrophysical $s$- and $r$-processes, about which, great uncertainties and many open questions still exist \cite{arconesWhitePaperNuclear2017,mumpowerImpactIndividualNuclear2016,schatzTrendsNuclearAstrophysics2016,arnouldAstronuclearPhysicsTale2020,sguazzinFirstMeasurementNeutronEmission2025}. These data are also crucial ingredients for simulations of advanced nuclear energy systems \cite{colonnaAdvancedNuclearEnergy2010,sunTransmutationLonglivedFission2022,huEfficientTransmutationLonglived2025}, the production of diagnostic or therapeutic radionuclides \cite{qaimNuclearDataProduction2017}, and a wide range of applications including nuclear waste management and innovative fuel cycles \cite{colonnaNeutronPhysicsAccelerators2018}. 

Yttrium provides a particularly instructive test bed for studies of both nuclear astrophysics and nuclear applications. Its only stable isotope ($^{89}$Y) locates at the $N=50$ shell closure. Together with its neighboring isotopes, $^{88}$Sr and $^{90}$Zr, $^{89}$Y contributes directly to the first $s$-process abundance peak \cite{taglienteHighresolutionCrossSection2024}. 
In nuclear astrophysical scenarios, \nuc{89}{Y} can be synthesized via neutron capture reaction on the unstable isotope \nuc{88}{Y} ($T_{1/2}=107$ days), or via proton capture reactions by \nuc{88}{Sr} following $\beta^+$ decays of \nuc{88}{Y}.
As a result, the nuclear flows involving \nuc{88}{Y} and \nuc{89}{Y} could affect the production of heavy elements up to the second $s$-process peak, which corresponds to the next bottleneck at $^{138}$Ba, $^{139}$La, and $^{140}$Ce \cite{taglienteHighresolutionCrossSection2024}. In addition, \nuc{88}{Y} is regarded as a suitable candidate to provide an observable signature of supernova neutrinos since its production can be affected by the neutrino-induced reaction \nuc{88}{Sr}($\nu_e$, $e^-$)\nuc{88}{Y} \cite{sieverdingNProcessLightImproved2018}. 
The $^{88}$Y($n,\gamma$)\nuc{89}{Y} reaction is also of great interest for reaction networks in the context of stockpile stewardship \cite{hoffman2006modeled}.

However, high radioactivity of $^{88}\mathrm{Y}$ makes direct measurements of the \nuc{88}{Y}$(n,\gamma)$ reaction cross sections extremely difficult. Experiments with inverse kinematics are also not feasible due to the lack of a free-neutron targets \cite{perezsanchezSimultaneousDeterminationNeutronInduced2020}. As a consequence, no experimental data on \nuc{88}{Y}$(n,\gamma)$ cross sections is available within the entire neutron energy range. For the above reasons, one has to rely on theoretical models to estimate the \nuc{88}{Y}$(n,\gamma)$ cross sections. Nevertheless, great discrepancies exist in theoretical predictions due to difficulties in describing the de-excitation process of the compound nucleus (CN), which depends on the knowledge of its $\gamma$-ray strength function ($\gamma$SF), nuclear level density (NLD) and its spin distributions, and the average radiative width $\langle\Gamma_{\gamma}\rangle$, for which different models give different predictions \cite{arnouldRprocessStellarNucleosynthesis2007, perezsanchezSimultaneousDeterminationNeutronInduced2020, sguazzinFirstMeasurementNeutronEmission2025}.
In 2016, Larsen $et \ al.$ \cite{larsenExperimentallyConstrained892016} predicted the $^{88}\mathrm{Y}(n,\gamma)$\nuc{89}{Y} cross section by using the Oslo method, with which both the $\gamma$SF and NLD are extracted from experimental data of the $^{89}\mathrm{Y}(p,p'\gamma)$\nuc{89}{Y} reaction. Due to uncertainties in the above quantities, the resulting \nuc{88}{Y}$(n,\gamma)$ reaction cross sections inherit rather large uncertainties \cite{larsenExperimentallyConstrained892016}. Consequently, the astrophysical reaction rates derived from these cross sections also carry large uncertainties, and show a significant deviation from the recommended rates in the JINA REACLIB database \cite{Cyburt_2010}.

The surrogate reaction method (SRM) is an indirect approach for determining nuclear reaction cross sections that cannot be measured directly \cite{escherCompoundnuclearReactionCross2012}. This method employs a more experimentally accessible ``target-projectile combination'' to produce the desired CN. It was initially used to infer cross sections of neutron-induced fission reactions \cite{brittSimulatedNfCross1979} and was later extended to neutron capture reactions, in which the Weisskopf-Ewing (WE) approximation was often employed \cite{younesNeutroninducedFissionCross2003,kessedjianNeutroninducedFissionCross2010,hughesUtilizingReactionsObtain2012}. The WE approximation assumes that the decay probabilities of a CN in a surrogate reaction is identical to those in the neutron-induced reactions, regardless of its spin-parity (SP) distributions. However, when applied to infer the $(n,\gamma)$ cross sections, the SRM with the WE approximation failed in most of the previously studied cases \cite{escherCompoundnuclearReactionCross2012,boutouxStudySurrogatereactionMethod2012,escherCrossSectionsNeutron2010,potelEstablishingTheoryDeuteroninduced2015}. It is particularly pronounced for the Y and Zr isotopes, since they are spherical and have lower level densities compared to those of actinides and rare-earth nuclei. This makes them more sensitive to the SP distributions of their corresponding compound nuclei \cite{escherCompoundnuclearReactionCross2012,ota87Yng89902015,forssenDeterminingNeutronCapture2007}. Recently, the SRM taking into account the SP distributions has been investigated to address the possible mismatch in the SPs of the CN in the surrogate and the neutron-induced reactions \cite{escherConstrainingNeutronCapture2018,ratkiewiczNeutronCaptureExotic2019,perezsanchezSimultaneousDeterminationNeutronInduced2020,tanPa233Cross2024}. These studies span from medium-mass ($^{87}$Y, $^{90}$Zr and $^{95}$Mo) isotopes to heavy actinides ($^{233}$Pa and $^{239}$Pu), demonstrating the importance of taking into account the SP of the CNs in studying the $(n,\gamma)$ reactions with the SRM. In 2025, Sguazzin $et$ $al.$ \cite{sguazzinFirstMeasurementNeutronEmission2025} successfully performed the first surrogate-reaction experiment in inverse kinematics, using the $^{208}$Pb$(p,p')$ reaction as a surrogate for the \nuc{207}Pb$(n,\gamma)$ reaction at neutron energies above 800 keV. Yet the above studies still rely on a single surrogate observable (for example, the $\gamma$-decay probability) to constrain the $\gamma$SF and NLD that strongly affect the inference of the desired $(n,\gamma)$ cross sections. Considering the complex dependence on nuclear model parameters, additional experimental constraints are necessary to confine them for reliable predictions of $(n,\gamma)$ cross sections for radioactive nuclei.

In this letter, we demonstrate the extraction of the $^{88}\mathrm{Y}(n,\gamma)$ reaction cross sections using the $(p,p'\gamma)$ surrogate reaction. \cite{ratkiewiczNeutronCaptureExotic2019,ducasse2016investigation}, 
Different from many previous applications of the SRM, such as those in Refs. \cite{escherConstrainingNeutronCapture2018, ratkiewiczNeutronCaptureExotic2019,tanPa233Cross2024}, where only surrogate reaction data, such as the experimentally measured $\gamma$-decay probability, $P_{\gamma}$, were used to confine the nuclear model parameters, we make use of experimental data of both $P_{\gamma}$ and $\gamma$SF to constrain the parameters of NLD and $\gamma$SF. Such optimization is achieved with a joint optimization framework combining Bayesian optimization and Markov chain Monte Carlo method (BO-MCMC), which enables an efficient optimization in multi-dimensional nuclear parameter spaces. As a result, uncertainties of the extracted $^{88}\mathrm{Y}(n,\gamma)$ cross sections in work are reduced significantly. Before calculating the desired \nuc{88}{Y}$(n,\gamma)$ cross sections with these parameters, we firstly test them in calculating the \nuc{88}{Sr}$(p,\gamma)$ cross sections with the SRM, which has existing experimental data of direct measurement to verify these parameters. This provides us a good basis for using these parameters in calculating the \nuc{88}{Y}$(n,\gamma)$ cross sections. 

When an incident neutron fuses with a \nuc{88}{Y} nucleus, a compound nucleus \nuc{89}{Y} is formed and decays subsequently primarily through $\gamma$ emission, neutron emission. The \nuc{88}{Y}$(n,\gamma)$ reaction corresponds to the $\gamma$-decay processes. Thus, according to the Hauser-Feshbach (HF) statistical reaction formalism, the \nuc{88}{Y}$(n,\gamma)$ reaction cross section at an neutron incident energy $E_n$ is \cite{hauserInelasticScatteringNeutrons1952,escherCompoundnuclearReactionCross2012}:
\begin{equation}
    \begin{aligned}
        \sigma_{n\gamma}(E_n) = \sum_{J\pi} \sigma_n^{CN}(E^\ast, J, \pi) G_\gamma^{CN}(E^\ast, J, \pi),
        \label{eq:sigma}
    \end{aligned}
\end{equation}
where $\sigma_n^{CN}(E^\ast,J,\pi)$ is the cross section of forming \nuc{89}{Y} with excitation energy $E^\ast$ and SP $J^\pi$, 
and $G_{\gamma}^{CN}(E^\ast,J,\pi)$ is the corresponding $\gamma$ decay branching ratio. The neutron incident energy is related to $E^\ast$ via $E_n = (1+\frac{1}{A})(E^\ast-S_n)$, where $A$ is the mass number of the target nucleus and $S_n$ is the neutron separation energy in the ground state of the compound nucleus.
While $\sigma_n^{CN}(E^\ast,J,\pi)$ can be calculated with the optical model using appropriate nuclear optical model potentials (OMPs), such as the global OMPs in Refs. \cite{koningLocalGlobalNucleon2003, baugeLaneconsistentSemimicroscopicNucleonnucleus2001, gorielyIsovectorImaginaryNeutron2007}, $G_{\gamma}^{CN}(E^\ast,J,\pi)$ for the $n$+\nuc{88}{Y} system cannot be measured directly and has to be determined indirectly with, for instance, the surrogate reaction method used in this work. 

Under the assumption that the CN mainly decays through $\gamma$ emission and neutron emission, the $\gamma$ decay branching ratio is \cite{capoteRIPLReferenceInput2009}:
\begin{equation}
        G_{\gamma}^\textrm{CN}{(E^\ast,J,\pi)}=\frac{T_{\gamma}(E^\ast,J,\pi)}{T_{\gamma}(E^\ast,J,\pi )+T_{n}(E^\ast,J,\pi )},
        \label{eq:G}
\end{equation}
where $T_\gamma$ and $T_n$ are the transmission coefficients for $\gamma$ emission and neutron emission, respectively. The transmission coefficient of $\gamma$ decay can be further expressed as:
\begin{equation}\label{eq-T}
T_\gamma(E^\ast,J,\pi)=\sum_{f} \sum_{X\ell}2\pi (E_\gamma^f)^{2\ell+1} f_{X\ell}(E_\gamma^f),
\end{equation}
where $E_\gamma^f$ is the energy of the emitted $\gamma$-ray from the excited state of the CN (with an excitation energy $E^\ast$) to its final state $f$, $X=\textrm{E}$ or M for electric (E) or magnetic (M) radiations, and $\ell$ is the multipolarity of the transition, and $f_{X\ell}(E_\gamma^f)$ is the corresponding $\gamma$SF.

For the $(n,\gamma)$ reaction with an radioactive nucleus, the $\gamma$ decay branching ratio has to be calculated theoretically because it cannot be measured directly. The main model parameters are those for calculating the NLDs and $\gamma$SFs of the CN. In the SRM, these parameters are constrained with experimental data of $\gamma$-decay probability $P_\gamma$ of the CN, which relates to the desired $\gamma$ decay branching ratios as:
\begin{equation}
    \begin{aligned}
        P_{\gamma}(E^\ast) = \sum_{J\pi}F^\textrm{CN}_\delta(E^\ast,J,\pi)G^{CN}_\gamma(E^\ast,J,\pi),
        \label{eq:P}
    \end{aligned}
\end{equation}
where $F^\textrm{CN}_\delta(E^\ast,J,\pi)$ is the SP distribution of the CN formed in a surrogate reaction. In this work, proton inelastic scattering from a \nuc{89}{Y} target, which raises the \nuc{89}{Y} target to its excited states that are above its neutron emission threshold, is chosen as the surrogate reaction. The SP of the compound nucleus \nuc{89}{Y} can be obtained from the theoretically calculated $(p,p'\gamma)$ cross sections on \nuc{89}{Y} within the coupled-channel reaction framework. The SRM uses thus obtained $F^\textrm{CN}_\delta(E^\ast,J,\pi)$ values to determine the nuclear model parameters that are necessary to calculate the $\gamma$-decay branching ratios $G_{\gamma}^\textrm{CN}{(E^\ast,J,\pi)}$ and uses the latter to calculate the $(n,\gamma)$ reaction cross sections according to Eq. (\ref{eq:sigma}). This is accomplished by optimizing these parameters so that the calculated $\gamma$-decay probabilities according to Eq. (\ref{eq:P}) best reproduce the experimentally measured $P_{\gamma}(E^\ast)$ values. Unlike what has been made in most previous studies with the SRM, which only make use of the experimental data of $P_{\gamma}(E^\ast)$ to confine these model parameters, experimentally measured $\gamma$SF values are also used in this work. As we will see, uncertainties of the theoretically predicted \nuc{88}{Y}$(n,\gamma)$ reaction cross sections are considerably reduced when these additional experimental data are used in the parameter optimization.

\begin{figure}
\centering
\includegraphics[width=\columnwidth]{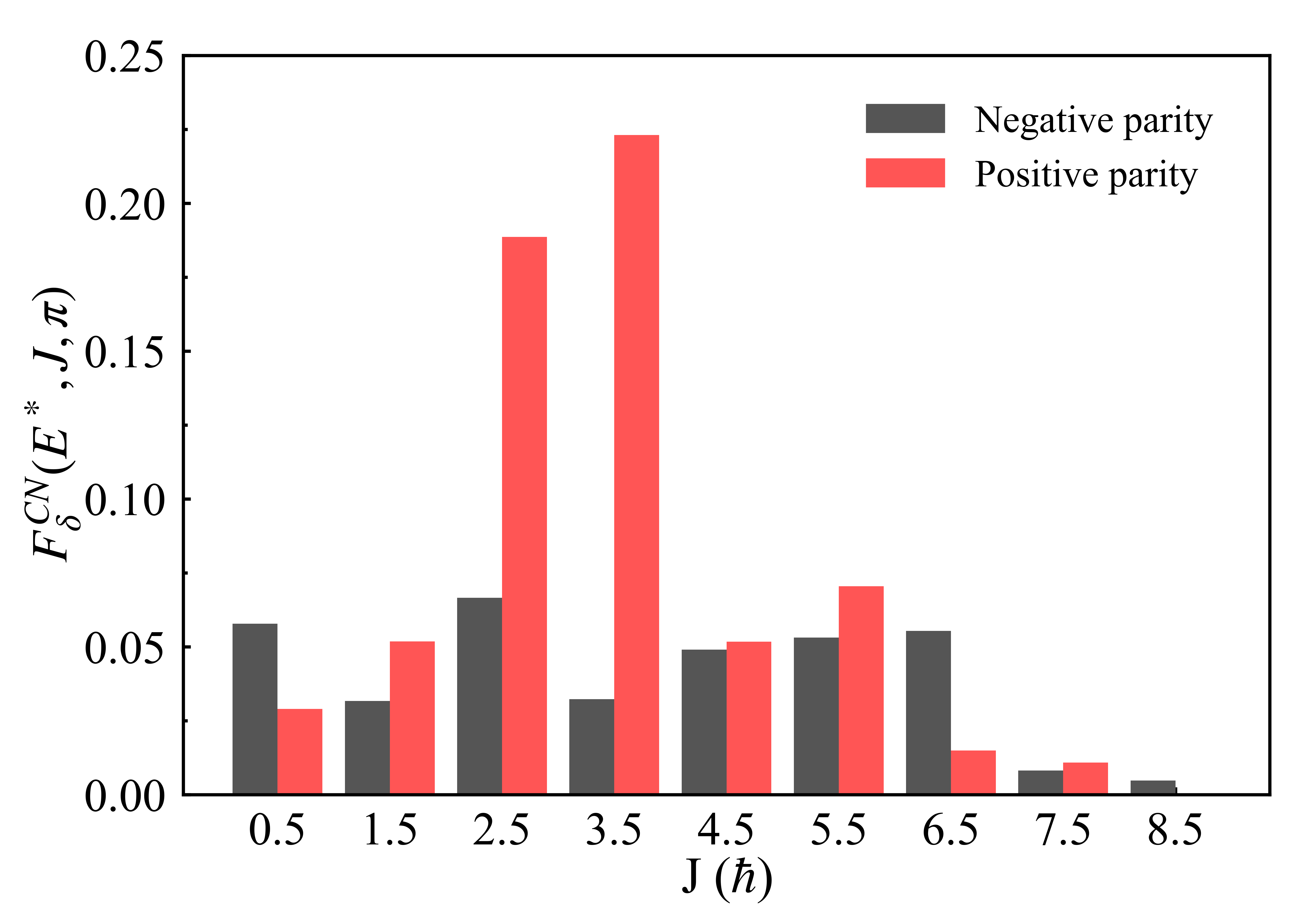}
\caption{Calculated SP distribution $F^{CN}_\delta(E^\ast,J,\pi)$ of $^{89}\mathrm{Y}^*$ populated by the $^{89}\mathrm{Y}(p,p'\gamma)$ at $E^\ast$=11.5 MeV, with red bars representing positive parities and black ones representing negative parities.}
\label{fig:F}
\end{figure}

--\textit{Spin parity distribution of the CN:}
To go beyond the WE approximation, we calculate the SP distribution, $F^{CN}_\delta(E^\ast,J,\pi)$, of $^{89}\mathrm{Y}^\ast$ populated by proton inelastic scattering on \nuc{89}{Y}. Cross sections corresponding to the excitation energies of \nuc{89}{Y} from 7.6 MeV to 13.3 MeV are obtained from coupled-channel calculations. Spin-parities of these excited states span from $1/2^{\pm}$ to $9/2^-$. These inelastic scattering channels are coupled with the ground and bound excited states of up to 3.99 MeV measured in Ref. \cite{awaya89Reaction14711967} assuming a rotor model for the target nucleus. Both electronic and magnetic transitions are included up to maximum multipolarity $\lambda=5$. Deformation parameters of these states are taken from Ref. \cite{awaya89Reaction14711967}, which are found to describe the low lying excited states data at 14.71 MeV incident energy therein satisfactorily with the optical model potentials taken from Ref. \cite{koningLocalGlobalNucleon2003}. The calculations are made with the computer code FRESCO \cite{thompsonCoupledReactionChannels1988}. The $F^{CN}_\delta(E^\ast,J,\pi)$ values from these calculations exhibits a slow variation within $E^\ast \in [7.6, 13.3]$, and the result near the neutron separation energy $S_n$=11.48 MeV is shown in \cref{fig:F}, which corresponds to an averaged spin $\bar{J}=3.4\ \hbar$. 
This value is slightly lower than the predicted value, $
\bar{J}=4.3\ \hbar$, with TALYS (version 2.0) \cite{koningTALYSModelingNuclear2023} for the CN produced in neutron capture at the same excitation energy. The coupled-channel calculations are made at a 22 MeV proton incident energy for getting the $F^{CN}_\delta(E^\ast,J,\pi)$ values because the corresponding $\gamma$-decay probability is measured at such an incident energy.

\begin{figure}
\centering
\includegraphics[width=\columnwidth]{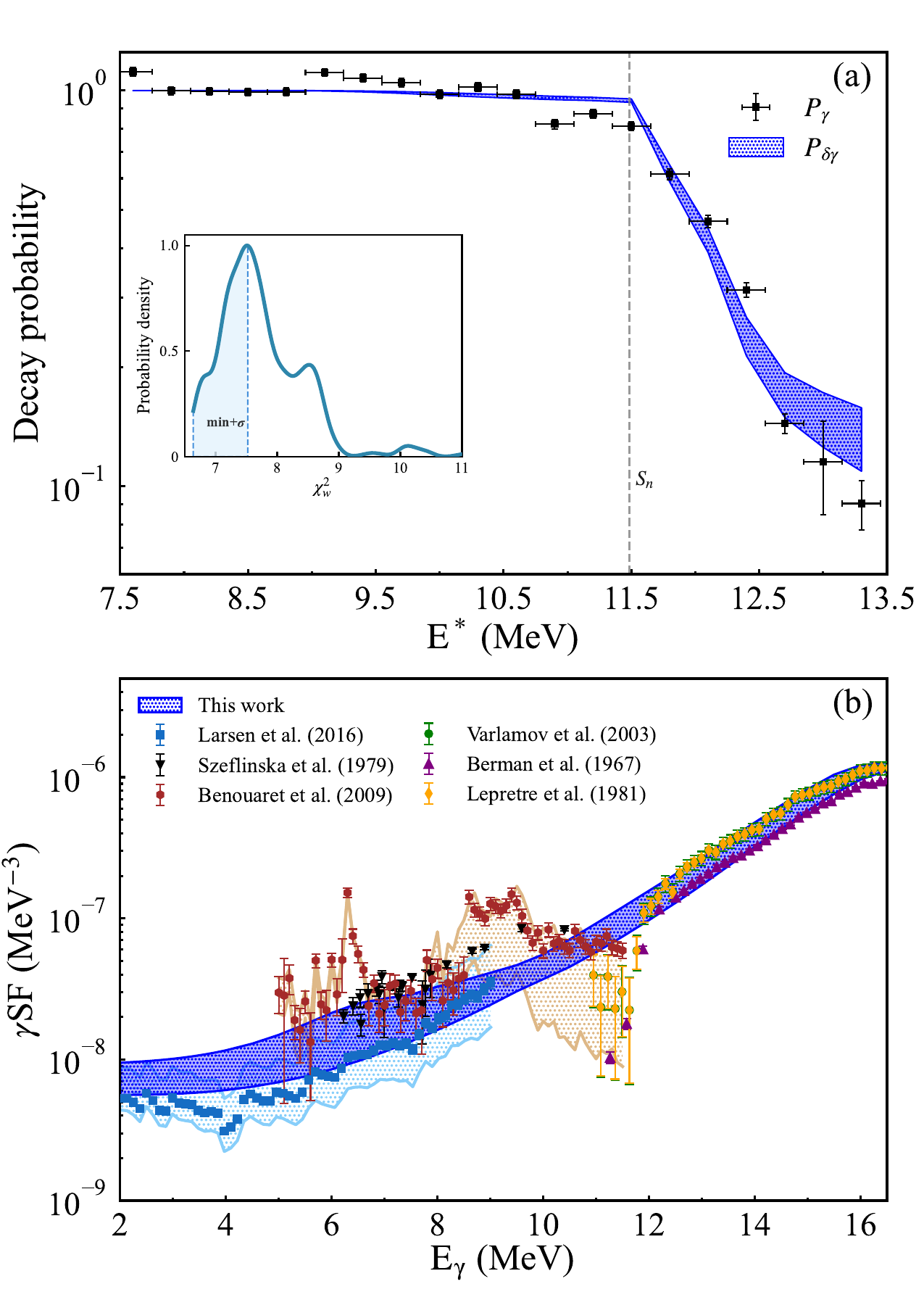}
\caption{(a) Decay probability for $\gamma$ emission (black dots) measured for the $^{89}\mathrm{Y}(p,p'\gamma)^{89}\mathrm{Y}^\ast$ reaction as a function of $E^\ast$ of $^{89}\mathrm{Y}^*$. The $P_{\gamma}(E^\ast)$ calculated with \cref{eq:P} using the adjusted TALYS parameters from the HCPS are shown in the blue shaded area. The vertical dotted line indicates the $S_n$ of $^{89}\mathrm{Y}$. The inset panel shows the probability density distribution of $\chi^2_{\text{w}}$ calculated with the BO-MCMC framework, where the blue shaded area represents the $\chi^2_{\text{w}}$ values corresponding to the HCPS. (b) The $\gamma$SF for $^{89}\mathrm{Y}$ as a function of $E_\gamma$. The blue shaded area represents the data fitted with the experimental \cite{larsenExperimentallyConstrained892016,benouaretDipoleStrength892009,szeflinskaGammarayStrengthFunctions1979,bermanPhotoneutronCrossSections1967,lepretreGiantDipoleStates1971} and evaluated \cite{varlamov2003photoneutron} data, where the same HCPS shown in panel (a) are employed.}
\label{fig:S+P}
\end{figure}

--\textit{$\gamma$-decay probability:}
The $\gamma$-decay probability data were collected using the HI-13 tandem accelerator at the China Institute of Atomic Energy \cite{T53-2024-0637,zhangSteppedupDevelopmentAccelerator2024}. An isotopically enriched \nuc{89}{Y} target with an areal density of $330\ \mu\mathrm{g}/\mathrm{cm}^2$ was bombarded by a proton beam of 1 nA and 22 MeV for 93 hours. The scattered protons were detected by two $\Delta E$-$E$ silicon telescopes positioned at $+65^\circ$ and $-100^\circ$ angles relative to the proton beam axis \cite{lesherSTARSLiBerACESegmented2010}. Each telescope consists of a $90$-$\mu\mathrm{m}$ double-sided silicon strip detector (DSSD) and two 1500-$\mu\mathrm{m}$ quad-silicon detectors (QSD) located 75 mm downstream of the target. A 0.5-$\mu\mathrm{m}$ Mylar foil was placed upstream of the DSSD to suppress the electron background. The silicon telescopes at $+65^\circ$ and $-100^\circ$ have detection efficiencies of 3.69\% and 2.61\%, respectively. Their energy resolutions range from 0.35\% to 0.65\%, which are sufficient for identification of light particles. Prompt $\gamma$ rays were detected by 25 HPGe coaxial detectors and eight Clover detectors surrounding the target chamber, each equipped with lead, cadmium and copper shieldings. The resulting energy resolution was determined to be 2.99 keV at 1.33 MeV. Therefore, the weighted average $P_{\gamma}(E^\ast)$ was obtained at the two angles by measuring $N_p$, the total number of detected protons, and $N_{\gamma}$, the number of coincidences between a proton and the $\gamma$ ray that identifies the relevant exit channel: $P_{\gamma}(E^\ast) = N_{\gamma}(E^\ast)/[N_p(E^\ast)\epsilon_\gamma]$. Here, $\epsilon_\gamma$ denotes the photopeak efficiency for detecting the exit channel $\gamma$ ray. The results are shown in \cref{fig:S+P}(a), where the error bars include statistical and systematic uncertainties, as well as the covariance between the measured quantities.

This is for the first time that the $P_{\gamma}(E^\ast)$ of the surrogate reaction forming the $^{89}\mathrm{Y}$ has been measured. \cref{fig:S+P}(a) illustrates that below $S_n$=11.48 MeV, only the $\gamma$-decay channel is open and the associated probability approaches unity. Above $S_n$, neutron emission starts to compete with $\gamma$ decay, and the sum of their probabilities must be unity. Our data satisfy this condition within the uncertainty limits of the measurements. This induces the trend that $P_{\gamma}(E^\ast)$ decreases with increasing excitation energy, as shown in \cref{fig:S+P}(a). As a result, it validates our experimental procedure and demonstrates the reliability of the measured $P_{\gamma}(E^\ast)$ for constraining the parameters of nuclear models. The emission of charged particles is massively hindered by the Coulomb barrier.

--\textit{Parameter optimization:}
To model the deexcitation of $^{89}\mathrm{Y}^\ast$, the following main ingredients are considered: the $\gamma$SF was determined with $E_1$ and $M_1$ components using generalized Lorentzian form \cite{kopeckyTestGammarayStrength1990} and the standard Lorentzian form description \cite{capoteRIPLReferenceInput2009}, respectively. The NLD was described with the Constant Temperature + Fermi gas model (CTM) \cite{gilbertCompositeNuclearlevelDensity1965}. In our case, eight key parameters in TALYS were employed to fit the experimental $P_{\gamma}(E^\ast)$ and both the experimental \cite{larsenExperimentallyConstrained892016,bermanPhotoneutronCrossSections1967} and evaluated \cite{varlamov2003photoneutron} $\gamma$SFs. These resonance parameters are energy $E$, strength $\sigma$, and width $\Gamma$ for the $\gamma$SF $E_1$ and $M_1$ components, the level density parameter $a$ and shell damping parameter $\gamma$ for the NLD.

In order to simultaneously constrain the parameters for calculating the $\gamma$-decay branching ratios and the $\gamma$-ray strength functions using both the experimental data $P_{\gamma}(E^\ast)$ and $\gamma$SF, we developed a BO-MCMC framework based on the Bayesian optimization (BO) and Markov chain Monte Carlo (MCMC) method. It adjusts the key parameters in TALYS for the calculations of $G_{\gamma}^{CN}(E^\ast,J,\pi)$, so that, together with the previously obtained $F_\delta^\textrm{CN}(E^\ast,J,\pi)$ values, the calculated $\gamma$-decay probability agree with it experimental value within the range of excitation energy $E^\ast\in [7.6,13.3]$ MeV. Simultaneously, the parameters for calculating the $\gamma$SFs are also found to reproduce the experimental $\gamma$SF values within the range of $E_\gamma\in[2,17]$ MeV.
The measured $P_\gamma(E^\ast)$ and $\gamma$SF values are reproduced by minimizing a weighted objective function: $\chi^2_{\text{w}} = 0.6 \cdot \chi^2(P_{\gamma}) + 0.4 \cdot \chi^2(\gamma \text{SF})$, where $\chi^2(P_{\gamma})$ and $\chi^2(\gamma \text{SF})$ represent the $\chi^2$ values that quantify the discrepancies between theoretical and experimental data for $P_{\gamma}(E^\ast)$ and $\gamma$SFs, respectively. The resulting optimized parameters are: for $\gamma$SF, $E_{E1}$=17.04 MeV, $\sigma_{E1}$=219.15 mb, $\Gamma_{E1}$=5.23 MeV, $E_{M1}$=7.82 MeV, $\sigma_{M1}$=0.75 mb and $\Gamma_{M1}$=2.61 MeV; and for the NLD, $a$=11.67 and $\gamma$=0.053 MeV, which correspond to a $\chi^2$-value $\chi_\textrm{w}^2=6.6$. The comparisons between calculated and experimetnal values of $P_\gamma(E^\ast)$ and $\gamma$SFs are shown in Fig. \ref{fig:S+P}.

The BO-MCMC optimization procedure is the following: 1) Three parallel Bayesian optimization algorithms individually vary the above parameters and compare the resulting $P_{\gamma}(E^\ast)$ and $\gamma$SFs to the experimental data to make initial coarse parameter estimation; 2) The top 5\% parameter sets with the lowest $\chi^2_{\text{w}}$ values from each of the three parallel BO runs are selected and combined to identify the overlapping parameter space, defining a confidence parameter region; 3) MCMC samplings within such a region are performed assuming uniform parameter distributions, leading to tens of thousands of calculations of both $P_{\gamma}(E^\ast)$ and $\gamma$SFs and yielding optimal parameter values that minimize the weighted $\chi^2_{\text{w}}$ function; 4) From all MCMC-sampled parameter sets, those satisfying $\chi^2_{\text{w}} \leq \min(\chi^2_{\text{w}}) + \mathrm{std}(\chi^2_{\text{w}})$ are selected as high confidence parameter sets (HCPS). The $\gamma$-decay branching ratios are then calculated with the HCPS before calculating the $^{88}\mathrm{Y}(n,\gamma)$ cross sections according to \cref{eq:sigma}. Uncertainties of these cross sections are determined by the size of the HCPS, which are defined as $(\sigma_\textrm{max}-\sigma_\textrm{min})/2\langle\sigma\rangle$, where $\sigma_\textrm{max}$, $\sigma_\textrm{min}$, and $\langle\sigma\rangle$ are respectively the maximum, minimum, and mean values of the $(n,\gamma)$ cross sections calculated within the range of the HCPS.

\begin{figure}
\centering
\includegraphics[width=\columnwidth]{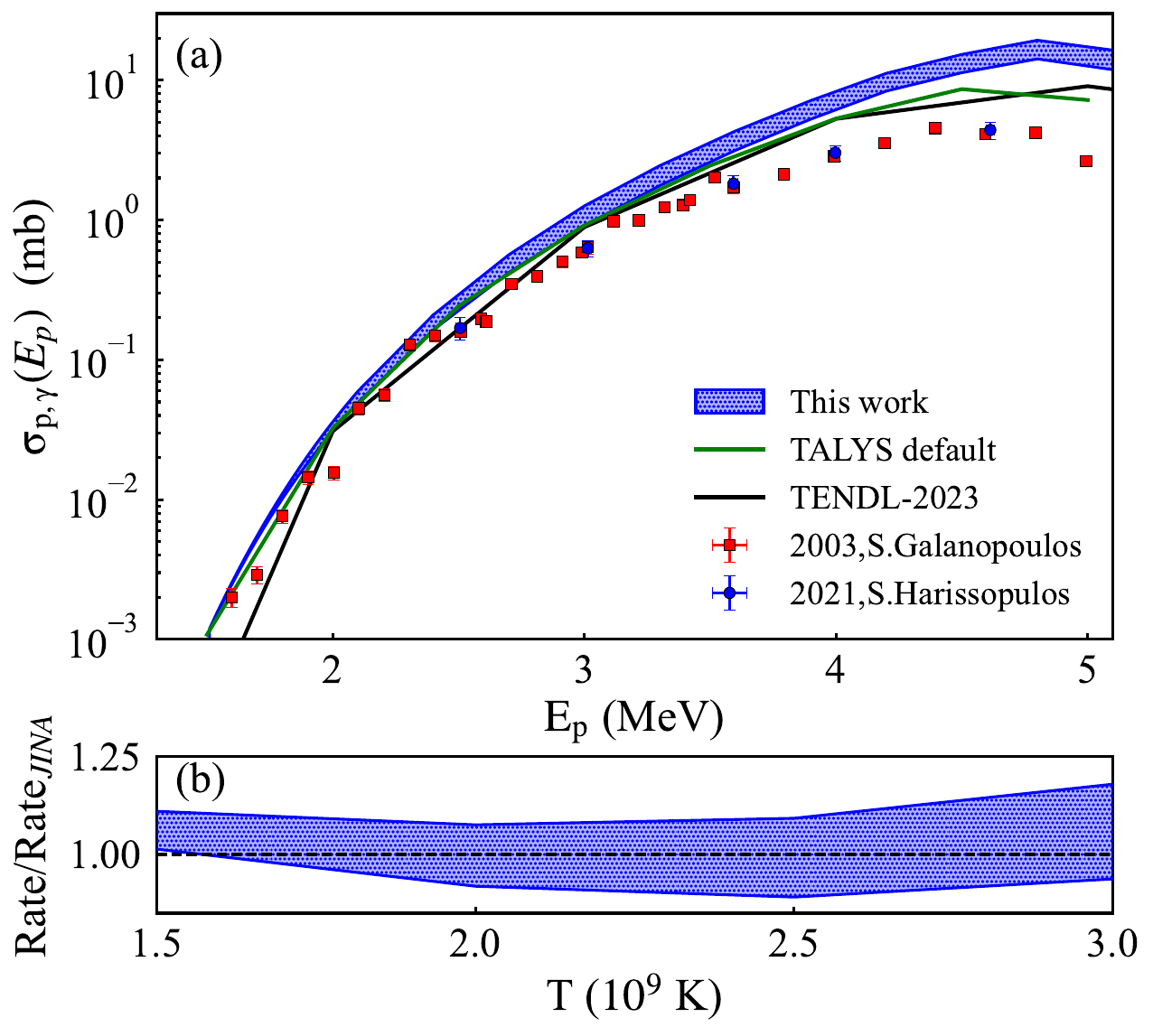}
\caption{(a) \nuc{88}{Sr}(p,$\gamma$) cross section as a function of proton energy. Our results (blue shaded area) are compared with experimental data from Harissopulos $et \ al.$ \cite{harissopulosCrossSectionMeasurements2021} and Galanopoulos $et \ al.$ \cite{galanopoulos88Sr892003}, as well as with the TALYS default calculations and TENDL-2023. (b) Ratio to JINA REACLIB evaluation for \nuc{88}{Sr}(p,$\gamma$) reaction rates as a function of astrophysical temperature.}
\label{fig:Sr_sigma_rate}
\end{figure}

--\textit{Validation of the model parameters:}
Before calculating the \nuc{88}{Y}($n,\gamma$) cross sections using these parameters in the framework of the SRM, we firstly test them in the calculation of the \nuc{88}{Sr}($p,\gamma$) cross section. Both reactions have the same CN, \nuc{89}{Y}, and can be inferred with the SRM, but the latter has directly measured experimental data to be compared with. The results are shown in \cref{fig:Sr_sigma_rate}, where \cref{fig:Sr_sigma_rate}(a) shows that the inferred \nuc{88}{Sr}($p$,$\gamma$) data are in line with the experimental ones below proton incident energy $E_p=3.5$ MeV, which is the most important energy range for nuclear astrophysical studies. Fig. \ref{fig:Sr_sigma_rate}(b) shows the ratios of \nuc{88}{Sr}($p$,$\gamma$) reaction rates relative to the $ths8$ rate compiled in the JINA REACLIB \cite{Cyburt_2010}, which also demonstrates an excellent consistency. These results suggest that our approach, which employs the $(p,p'\gamma)$ as a surrogate reaction and constraint the nuclear model parameters by simultaneously reproducing the experimental values of $P_\gamma(E^\ast)$ and $\gamma$SF is reliable for the extraction of the \nuc{88}{Y}$(n,\gamma)$ cross sections.

\begin{figure}
\centering
\includegraphics[width=\columnwidth]{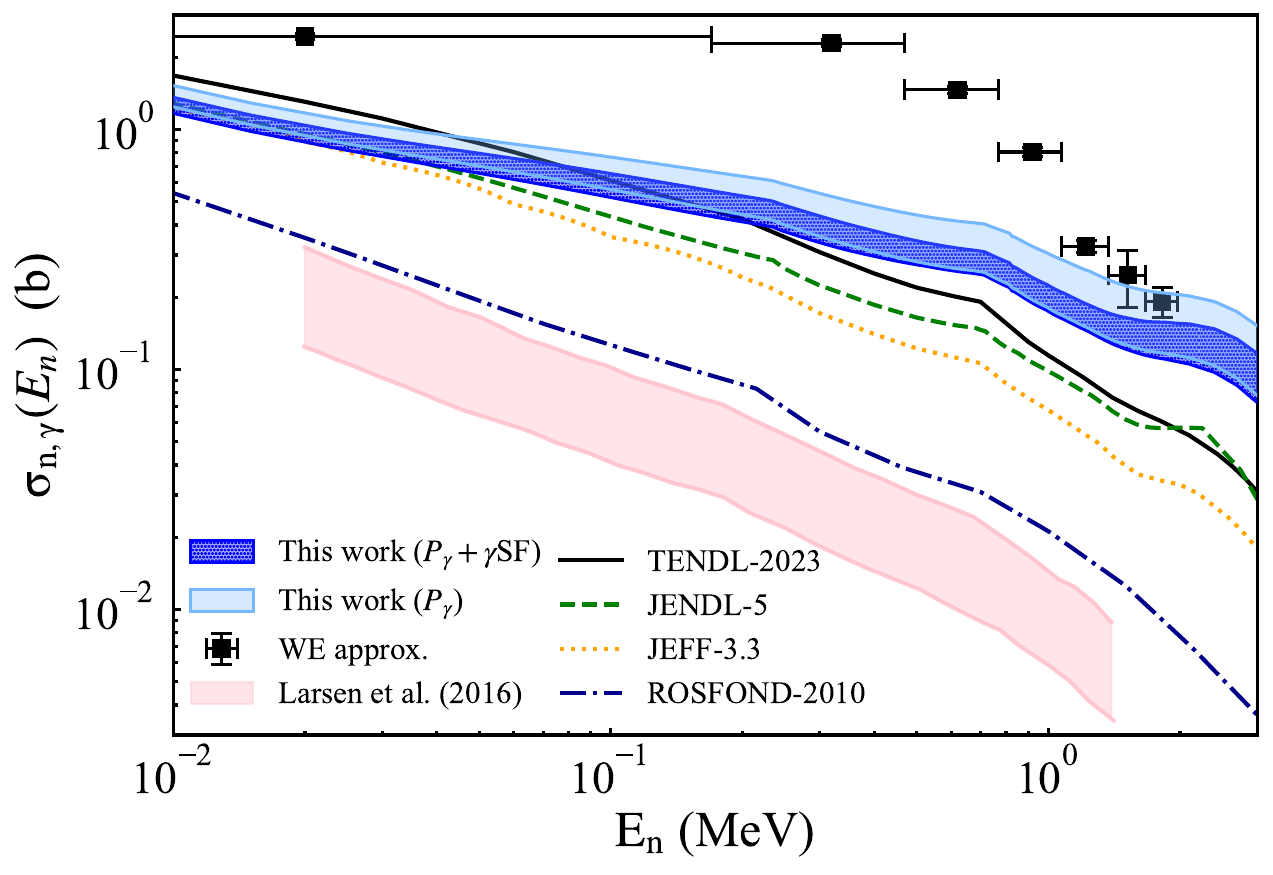}
\caption{$^{88}\mathrm{Y}(n,\gamma)$ cross sections as a function of neutron energy. The blue shaded area shows the data from joint constraints of both $P_{\gamma}(E^\ast)$ and $\gamma$SF, while the light blue one indicates the data induced by only constraint of $P_{\gamma}(E^\ast)$. Here, the Jeukenne-Lejeune-Mahaux (JLM) microscopic OMP \cite{baugeLaneconsistentSemimicroscopicNucleonnucleus2001} and the CTM of the NLD are employed to interpret our data, and a width-fluctuation correction \cite{koningTALYSModelingNuclear2023} is taken into account by default. The black squares are the data obtained with the WE approximation. The pink shaded area represents the inferred data of A.C. Larsen $et \ al.$ \cite{larsenExperimentallyConstrained892016}. The black solid, green dashed, yellow dotted and blue dash-dotted lines indicate the available evaluations \cite{koningTENDLCompleteNuclear2019,iwamotoJapaneseEvaluatedNuclear2023,plompenJointEvaluatedFission2020,zabrodskaya2007rosfond}.}
\label{fig:sigma_N1}
\end{figure}

--\textit{The \nuc{88}{Y}$(n,\gamma)$ cross sections with the SRM:}
The $^{88}\mathrm{Y}(n,\gamma)$ cross sections predicted with the SRM are shown in \cref{fig:sigma_N1}. The uncertainties of these cross sections are between 7.6\% at neutron incident energy $E_{n}$=0.01 MeV and 23.1\% at $E_{n}$=3.0 MeV. Such uncertainties will increase by a factor of 1.8 when only the experimental data of $P_{\gamma}(E^\ast)$ were used to constrain the model parameters. Apparently, simultaneous fitting both $P_{\gamma}(E^\ast)$ and $\gamma$SF data, as mentioned above, provides a much stronger constraint on the model parameters, which allows us to get much smaller uncertainties in the SRM predicted cross sections.

At lower incident energies, $E_\textrm{n}\lesssim 0.06$ MeV, our result agree well with the evaluated data in TENDL-2023 \cite{koningTENDLCompleteNuclear2019}, JENDL-5 \cite{iwamotoJapaneseEvaluatedNuclear2023}, and JEFF-3.3 \cite{plompenJointEvaluatedFission2020}, but strongly disagree with the evaluated data in ROSFOND-2010 \cite{zabrodskaya2007rosfond}, the cross sections obtained with the Oslo method in Ref. \cite{larsenExperimentallyConstrained892016}, and the results obtained with the SRM but using the WE approximation. At higher incident energies, our derived $^{88}\mathrm{Y}(n,\gamma)$ cross sections disagree with the evaluated data in TENDL-2023, JENDL-5, and JEFF-3.3, and such discrepancies increase as $E_\textrm{n}$ increases. But, at the same time, our results begin to agree better with those derived with the WE approximation. Note that the same observation that the SRM predicted cross sections taking into account the SPs differ from those with the WE approximations more at low incident energies than at higher incident energies were also seen in previous studies (see, for instance, Ref. \cite{perezsanchezSimultaneousDeterminationNeutronInduced2020}). Clearly, our results suggest that it is crucial to take into account the SPs of the compound nuclei in deriving $(n,\gamma)$ reaction cross sections with the SRM. Together with constraining the nuclear model parameters by using experimental data of not only $P_\gamma(E^\ast)$ but also $\gamma$SFs, we have got the most accurate estimation of the $^{88}\mathrm{Y}(n,\gamma)$ cross sections for the first time.

\begin{figure}
\centering
\includegraphics[width=\columnwidth]{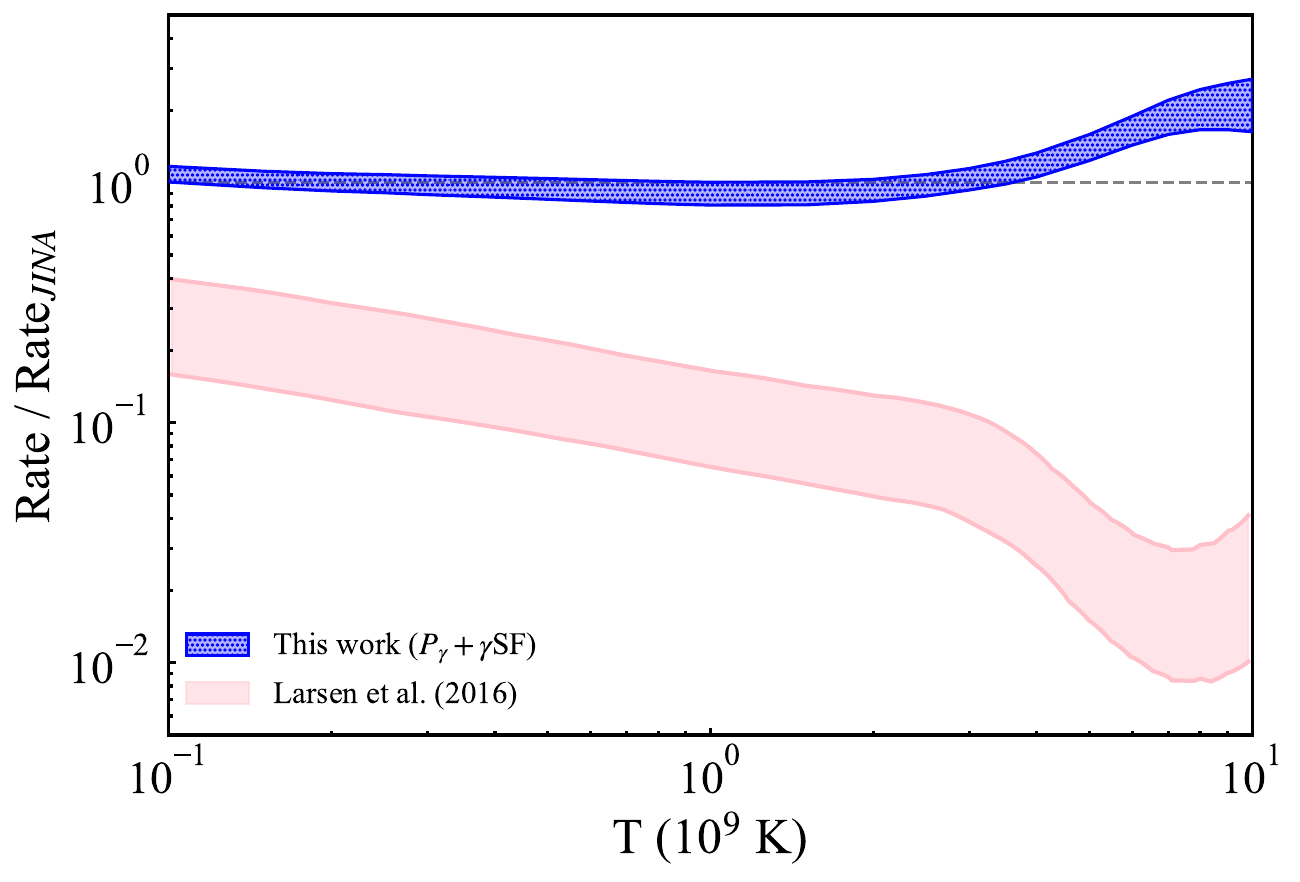}
\caption{Ratio to JINA REACLIB evaluation for \nuc{88}{Y}$(n,\gamma)$ reaction rates as a function of astrophysical temperature. The blue shaded area shows the data obtained from joint constraints of both $P_{\gamma}(E^\ast)$ and $\gamma$SF. The pink shaded area shows the calculations by Larsen $et \ al.$ \cite{larsenExperimentallyConstrained892016}.}
\label{fig:rate_JLM}
\end{figure}

The $^{88}\mathrm{Y}(n,\gamma)$ cross sections determined above allow us to calculate the $^{88}\mathrm{Y}(n,\gamma)$ astrophysical reaction rates. The results are shown in \cref{fig:rate_JLM} as the ratios between the reaction rates obtained in this work and those with the recommended label \textit{ths8} from JINA REACLIB database \cite{Cyburt_2010}. The same ratios are also shown for the $^{88}\mathrm{Y}(n,\gamma)$ cross sections determined with the Oslo method in Ref. \cite{larsenExperimentallyConstrained892016}. One sees that our results agree very well with the \textit{ths8} JINA REACLIB database at temperatures below around $4\times 10^9$ K, while results with the Oslo method underestimated the REACLIB data considerably.

\begin{figure}
\centering
\includegraphics[width=\columnwidth]{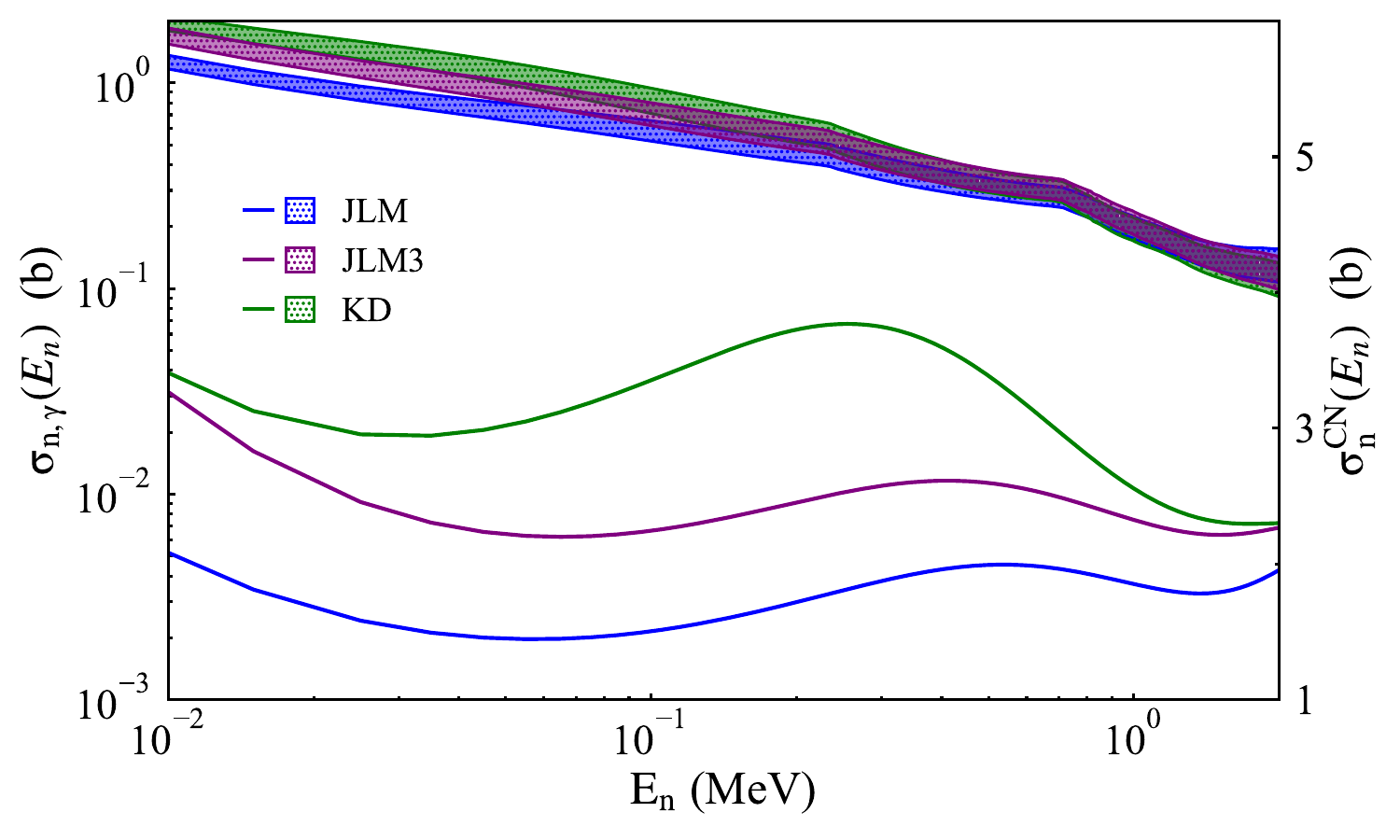}
\caption{$\sigma_n^{CN}(E_n)$ for forming $^{89}\mathrm{Y}$ (right-hand axis) and the resulting $^{88}\mathrm{Y}(n,\gamma)$ data (left-hand axis) as function of neutron energy, obtained with three different OMPs.}
\label{fig:sigmaCN+OMP}
\end{figure}

The OMPs play an important role in the calculation of $\sigma_n^{CN}(E_n)$, which directly affects the extraction of the neutron capture cross sections according to \cref{eq:sigma}. Let us discuss the influence of OMPs on the inference of $^{88}\mathrm{Y}(n,\gamma)$ data. TALYS 2.0 provides several commonly used OMPs, such as JLM \cite{baugeLaneconsistentSemimicroscopicNucleonnucleus2001}, a modified version of JLM with normalization for the imaginary potential of the OMP (case 3) (designated as JLM3) \cite{gorielyIsovectorImaginaryNeutron2007}, and KD \cite{koningLocalGlobalNucleon2003}. As shown in \cref{fig:sigmaCN+OMP}, three OMPs can result in very different curves for $\sigma_n^{CN}(E_n)$ for forming $^{89}\mathrm{Y}$, particularly at $E_{n} < 1$ MeV. However, these OMPs can hardly change the patterns of the deduced $^{88}\mathrm{Y}(n,\gamma)$ cross sections. This suggests that the inference of $^{88}\mathrm{Y}(n,\gamma)$ cross sections is not sensitive to the OMPs.

As mentioned above, the SP mismatch between the surrogate and desired reactions has been considered to infer the $^{88}\mathrm{Y}(n,\gamma)$ cross sections in a reasonable way. To further account for uncertainties in the calculated $F^{CN}_\delta(E^\ast,J,\pi)$ (see \cref{fig:F}), we vary the weights schematically by shifting the overall distribution by $\pm\hbar$, by which four different NLD models are employed to deepen our theoretical sensitivity studies. Among them, CTM \cite{gilbertCompositeNuclearlevelDensity1965} and Back-shifted Fermi Gas Model (BFM) \cite{dilgLevelDensityParameters1973} are phenomenological models, and Skyrme-Hartree-Fock-Bogolyubov level densities from numerical tables (SHFB) \cite{gorielyHARTREEFOCKNUCLEAR2001} and Temperature-dependent Gogny-Hartree-Fock-Bogolyubov combinatorial level densities from numerical tables (TGHFB) \cite{hilaireTemperaturedependentCombinatorialLevel2012} are microscopic NLD descriptions. The data showing model sensitivity are tabulated in \cref{tab:spin_sensitivity}. When shifting the $F^{CN}_\delta(E^\ast,J,\pi)$ distribution by $+\hbar$ (right) and $-\hbar$ (left), the resulting uncertainty of the $^{88}\mathrm{Y}(n,\gamma)$ data induced by CTM decreases by a factor of 0.11 and 0.19, respectively. Similar variation is also shown in the case of SHFB. Conversely, the TGHFB-induced $^{88}\mathrm{Y}(n,\gamma)$ data have an increased uncertainty for both right- and left-shifting $F^{CN}_\delta(E^\ast,J,\pi)$ distributions. In addition, for the right shift, the $^{88}\mathrm{Y}$ data uncertainty provided by the BFM description become visibly enlarged, whereas the left shift only results in a slight variation of the data uncertainty. Overall, the shifted $F^{CN}_\delta(E^\ast,J,\pi)$ distribution could result in a visible variation in the $^{88}\mathrm{Y}$ data uncertainty, suggesting that $^{89}$Y is sensitive to difference in the SP distribution. This is mainly due to the fact that $^{89}$Y is more spherical and has lower level densities compared to minor actinides and rare-earth nuclei \cite{escherCompoundnuclearReactionCross2012,ota87Yng89902015}.

\begin{table}
\centering
\caption{The variation of the $^{88}\mathrm{Y}(n,\gamma)$ data uncertainty caused by left- and right-shifting $F^{CN}_\delta(E^\ast,J,\pi)$ distribution under four different NLD models. The variation percentage represents the change in data uncertainty relative to the one from the initial distribution, as shown in \cref{fig:sigma_N1}.}
\label{tab:spin_sensitivity}
\begin{tabular*}{0.9\columnwidth}{@{\extracolsep{\fill}}c|cc@{}}
\hline
\multirow{2}{*}{NLD model} & \multicolumn{2}{c}{Variation percentage} \\
                     & $+1 \hbar$    & $-1\hbar$   \\
\hline
CTM                  & -10.6\%      & -19.4\%    \\
BFM                  & +30.1\%      & -6.5\%    \\
SHFB                 & -16.0\%     & -12.8\%    \\
TGHFB                & +49.5\%     & +21.6\%    \\
\hline
\end{tabular*}
\end{table}

In summary, we have presented that a measurement of the $^{89}\mathrm{Y}(p,p^\prime\gamma)$ reaction, when combined with proper theoretical treatment, can be used to indirectly and reliably determine the $^{88}\mathrm{Y}(n,\gamma)$ cross sections. The benchmark study of the \nuc{88}{Sr}($p,\gamma$) reaction illustrates that the $(p,p^\prime)$ inelastic scattering can be regarded as a surrogate reaction for neutron captures on unstable nuclei. Reliable $^{88}\mathrm{Y}(n,\gamma)$ cross sections are then extracted for the first time with uncertainties ranging from 7.6\% at $E_n$=0.01 MeV to 23.1\% at $E_n$=3.0 MeV, which helps to resolve the visible discrepancy among existing evaluated data. Moreover, the corresponding astrophysical reaction rates are found to be in excellent agreement with the JINA REACLIB recommendations. The above results shown that simultaneous constraints on $P_\gamma$ and $\gamma$SF data are useful for obtaining reliable $(n,\gamma)$ cross sections with the framework of SRM. Furthermore, our data inferred seem to be insensitive to the choice of OMPs, but sensitive to the SP distribution of CN. The present approach will open up perspectives on determining unknown $(n,\gamma)$ cross sections, with far-reaching implications for improving our understanding of heavy-element nucleosynthesis.

This work was supported by the the National Key R\&D Program of China (Grant Nos.: 2022YFA1603300, 2023YFA1606702, 2023YFA1606603), the National Natural Science Foundation of China (Grant Nos.: 12575133, 12235013, 11875214), the Stable Support for Basic Research Program (Grant No.: BJ010261223282) and the Nuclear Data Key Laboratory Foundation (Grant No.: JCKY2024201C151).

\bibliographystyle{apsrev4-2}
\bibliography{file}

\end{document}